\documentclass[prresearch,showpacs,twocolumn,amsmath,amssymb,floatfix,superscriptaddress]{revtex4-2}
\usepackage{graphicx} 
\usepackage{bm}
\usepackage{bbm}
\usepackage{color}
\usepackage{amsmath}
\usepackage{amssymb}
\usepackage{color}
\usepackage{footnote}
\usepackage{comment}
\usepackage{hyperref}
\usepackage{subfigure}
\usepackage{appendix}
\usepackage[applemac]{inputenc}

\begin{document}
\title{Nonlocal Josephson diode effect  in minimal Kitaev chains}
\author{Jorge Cayao}
\email{jorge.cayao@physics.uu.se}
\affiliation{Department of Physics and Astronomy, Uppsala University, Box 516, S-751 20 Uppsala, Sweden}
\author{Masatoshi Sato}
\email{msato@yukawa.kyoto-u.ac.jp}
\affiliation{Center for Gravitational Physics and Quantum Information, Yukawa Institute for Theoretical Physics, Kyoto University, Kyoto 606-8502, Japan}

\date{\today} 
\begin{abstract}
We study the emergence of the nonlocal Josephson effect in a system composed of three laterally coupled minimal Kitaev chains and exploit it to realize the nonlocal Josephson diode effect. We find that an imbalance between crossed Andreev reflections and electron cotunneling in the middle Kitaev chain gives rise to an asymmetric $2\pi$-periodic phase-dependent Andreev spectrum, controlled by the superconducting phases across the left and right junctions. We then show that the asymmetric Andreev spectrum, formed by hybridized Andreev bound states at the left and right junctions, enables a supercurrent across one junction via the phase difference at the other junction, thereby signaling the nonlocal Josephson effect. Notably, these nonlocal Josephson supercurrents exhibit distinct positive and negative critical currents, demonstrating the realization of the nonlocal Josephson diode effect with highly tunable polarity and efficiencies exceeding 50\%.  The nonlocal Josephson diode effect requires breaking local time-reversal and local charge-conjugation symmetries, with the latter being unique to minimal Kitaev chains. Our results establish minimal Kitaev chains as a highly controllable platform for engineering nonlocal Josephson phenomena.
\end{abstract}
\maketitle

\section{Introduction}
Nonlocality is an inherent quantum effect revealing instantaneous action at a distance, tied to the wavefunction nature, and has led to the prediction of several phenomena relevant to quantum technologies \cite{RevModPhys.86.419}. While mostly explored in relation to entanglement \cite{RevModPhys.82.665,cavalcanti2011quantum,villegas2024nonlocality}, the concept of nonlocality has been often studied in superconducting devices. Perhaps the simplest example is a two-terminal superconducting junction, where variations of voltages on one terminal give rise to a differential current on the second terminal, thus determining the so-called nonlocal conductance \cite{PhysRevB.106.L241301,PhysRevLett.124.036801,feng2025long,zhang2019next,frolov2019quest}. More recently, nonlocality was studied in multi-terminal Josephson junctions, where a supercurrent flows across one part of the system due to variations in the superconducting phases elsewhere, defining what is known as the nonlocal Josephson effect (JE) \cite{Pillet_2019}. Since Josephson junctions are key for superconducting circuits and computing \cite{devoret2004superconducting,clarke2008SC,Brecht2016,krantz2019quantum,martinis2020quantum,aguado2020perspective,benito2020hybrid,aguado2020majorana,PRXQuantum.2.040204,siddiqi2021engineering,bargerbos2022singlet,doi:10.1146/annurev-conmatphys-031119-050605,pita2023direct,tanakaReview2024,FukayaJPCM2025}, the nonlocal JE is directly relevant for quantum applications while offering a way to further delve the foundations of nonlocality.

The research on nonlocal JE has not only attracted theoretical efforts \cite{Pillet_2019,10.21468/SciPostPhysCore.2.2.009,PhysRevB.97.035443,PhysRevB.102.245435,PhysRevB.104.075402,PhysRevB.109.205406,PhysRevResearch.5.033199,PRXQuantum.5.020340,PhysRevB.109.245133,PhysRevB.110.235426,10.21468/SciPostPhys.17.2.037,PhysRevB.111.024506,mondal2026AMNNJE} but also its realization has already been reported in many experiments \cite{strambini2016omega,draelos2019supercurrent,PhysRevX.10.031051,graziano2022selective,Matsuo_2022,matsuo2023phase,Haxell_2023}. Most of the studies involved conventional spin-singlet $s$-wave superconductors, leaving unconventional superconductors seldom explored in spite of their intrinsic nonlocal order parameters.  One of the simplest systems that hosts controllable unconventional superconductivity with a $p$-wave order parameter can be engineered by connecting two quantum dots (QDs) via a superconductor \cite{PhysRevB.86.134528,PhysRevB.90.220501}.   This setup represents a minimal Kitaev chain, where electron cotunneling (ECT) and crossed Andreev reflections (CARs) mediate superconducting and normal processes between QDs  \cite{PhysRevB.86.134528,sau2012realizing}. Minimal Kitaev chains have lately triggered multiple efforts \cite{PhysRevB.106.L201404,PhysRevResearch.5.043182,PRXQuantum.5.010323,cayao2024pairPMMMs,liu2024enhancing,PhysRevB.110.245144,PhysRevB.109.035415,PhysRevB.110.165404,PhysRevB.110.245412,nitsch2024tetron,kotetes2024nonRecifourpi,cayao2024NHtwositeKitaev,PhysRevB.111.115419,vimal2025EntMKC} and their implementation has already been shown in superconductor-semiconductor hybrids \cite{dvir2023realization,bordin2023crossed,PhysRevX.13.031031,zatelli2023robustpoorMajo}.   Despite the significant attention, the nonlocal JE and its impact on minimal Kitaev chains have not yet been studied.

\begin{figure}[!b]
\centering
\includegraphics[width=0.45\textwidth]{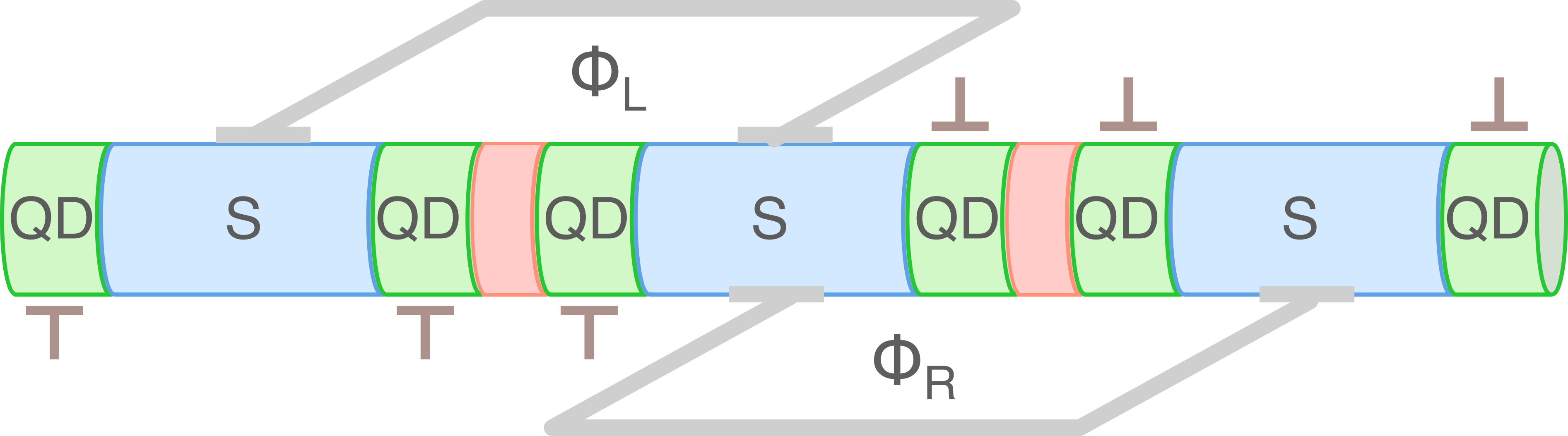}
\caption{Three   minimal Kitaev chains laterally coupled to form a double Josephson junction, where the phase of the left (right) junction is controlled by an external magnetic flux $\Phi_{\rm L (R)}$. Here, a minimal Kitaev chain is formed by connecting two quantum dots (QDs, green) by a superconductor  (S, blue). The onsite energy of the QDs is controlled by gates (bars, brown), while the coupling between minimal Kitaev chains can be tuned by an insulating barrier (red).}
\label{Fig0} 
\end{figure}

In this work, we consider three laterally coupled minimal Kitaev chains [Fig.\,\ref{Fig0}] and study the emergence of nonlocal Josephson phenomena. In particular, we demonstrate that, when  CAR and ECT   in the middle minimal Kitaev chain are distinct, an asymmetric phase-dependent Andreev spectrum appears and is controllable by the superconducting phases across the left and right junctions. We find that this asymmetric Andreev spectrum arises from hybridized Andreev bound states at the left and right junctions  and requires breaking local time-reversal and local charge-conjugation symmetries. Notably, the asymmetric Andreev spectrum induces a finite supercurrent across one junction by tuning the superconducting phase difference across the other junction, hence unveiling the nonlocal JE. Even more interesting is that the nonlocal supercurrents develop distinct positive and negative critical currents, giving rise to a nonreciprocal Josephson transport that defines the emergence of a nonlocal Josephson diode effect. We show that this diode effect can achieve robust efficiencies surpassing $50\%$ by controlling CAR and ECT in the middle chain and the phase difference in the second junction. Our findings put forward minimal Kitaev chains for realizing highly tunable nonlocal Josephson phenomena.


\section{JJs formed by three minimal Kitaev chains}
We consider a Josephson system where three minimal Kitaev chains described by
\begin{equation}
\begin{split}
H_{\alpha}(\phi_\alpha)&=\varepsilon_{\alpha_1}c^{\dagger}_{\alpha_1}c_{\alpha_1}+\varepsilon_{\alpha_2}c^{\dagger}_{\alpha_2}c_{\alpha_2}
\\
&+t_{\alpha}c_{\alpha_1}^{\dagger}c_{\alpha_2}
+\Delta_{\alpha} {\rm e}^{i\phi_{\rm \alpha}}c_{\alpha_1}^{\dagger}c_{\alpha_2}^{\dagger}+{\rm h. c.}\  
\end{split}    
\nonumber
\end{equation}
are laterally coupled  [Fig.\,\ref{Fig0}] and modelled by
\begin{equation}
\begin{split}
H(\phi_{\rm L}, \phi_R)
=\sum_\alpha H_\alpha(\phi_\alpha)
+\tau_{\rm L}c^{\dagger}_{\rm L_{2}}c_{\rm M_{1}}+\tau_{\rm R}c^{\dagger}_{\rm M_{2}}c_{\rm R_{1}}+{\rm h. c.}.
\label{MKCH}   
\end{split}    
\end{equation}
Here $\alpha={\rm L, M, R}$ labels the left (L), middle (M), and right (R) minimal Kitaev chains formed by two sites $j=1,2$, and $c_{\alpha_{j}}$ ($c_{\alpha_{j}}^{\dagger}$) destroys (creates)  an electronic state at site $j=1,2$ in the chain $\alpha$. The first and  second terms in $H_\alpha(\phi_\alpha)$ describe onsite energies $\varepsilon_{\alpha_{j}}$ controlled by gates [brown bars in Fig.\,\ref{Fig0}]. The third term represents hopping between sites on the same chain, a characteristic of ECT processes. The fourth term is a spinless superconducting potential, with amplitude $\Delta_{\alpha}$  and  phase $\phi_{\alpha}$, describing CAR processes; note that the  pair potential characterizes the $p$-wave nature of superconductivity, as in the Kitaev chain and irrespective of topology \cite{sato2017topological,tanakaReview2024}.  We model the coupling between the three chains by the second and third terms in Eq.\,(\ref{MKCH}) 
with tunneling amplitudes $\tau_{\rm L(R)}$ across the left (right) junctions. 
Using a gauge symmetry, we also set the phase of the middle chain as  $\phi_{\rm M}=0$, leading to the junction depending on two superconducting phases, $H(\phi_{\rm L},\phi_{\rm R})$.
The phase differences $\phi_{\rm L}$ and $\phi_{\rm R}$ between chains can be externally controlled by magnetic fluxes [Fig.\,\ref{Fig0}], as in Refs.\,\cite{Matsuo_2022,Coraiola_2023,Haxell_2023}.
Despite the simplicity, two-site Kitaev chains have already been experimentally realized in QDs coupled by a superconductor-semiconductor system \cite{dvir2023realization,bordin2023crossed,PhysRevX.13.031031,zatelli2023robustpoorMajo,Haaf2024}, where the control of ECT and CAR was demonstrated   \cite{Wang_Nat2022,Wang_Natcom2023}. 
 As we show below, the system exhibits a unique phase dependence, enabling a nonlocal JE with nonreciprocal origin.


\section{Breaking and preservation of symmetries}
We begin by analyzing symmetries of the Josephson system in Eq.\,(\ref{MKCH}).  First of all, the quasi-particle energy derived by Eq.\,(\ref{MKCH}) forms a pair $(-E, E)$ because of particle-hole symmetry  of the Bogoliubov-de Gennes Hamiltonian.
While the Josephson system respects time-reversal symmetry (TRS),  
$\mathcal{T}H(\phi_{\rm L},\phi_{\rm R})\mathcal{T}^{-1}=H(-\phi_{\rm L},-\phi_{\rm R})$, where
the time-reversal anti-unitary operator ${\cal T}$ is given by ${\cal T}c_{\alpha_i}{\cal T}^{-1}=c_{\alpha_i}$, it does not have either $\mathcal{T}H(\phi_{\rm L},\phi_{\rm R})\mathcal{T}^{-1}= H(-\phi_{\rm L},\phi_{\rm R})$ or $\mathcal{T}H(\phi_{\rm L},\phi_{\rm R})\mathcal{T}^{-1}= H(\phi_{\rm L},-\phi_{\rm R})$, implying local TRS is broken either across the left or right junctions, respectively.  For  $\tau_{\rm L(R)}\neq0$, the left (right) junction also explicitly breaks the local spatial inversion symmetry, even when the entire system may be spatially inversion-symmetric. Thus, each junction satisfies the standard criterion for the Josephson diode effect \cite{PhysRevLett.128.037001,yuan2022supercurrent,he2024,PhysRevLett.131.096001,PhysRevB.109.L081405}. 

However, our system Eq.\,(\ref{MKCH}) has an additional symmetry constraint for the Josephson diode effect, which originates from a special symmetry at the  ``sweet spot" ($\varepsilon_\alpha=0$, $t_\alpha=\Delta_\alpha$), where the chains support quadratically protected Majorana states \cite{PhysRevB.86.134528}. In fact, for $\varepsilon_\alpha=0$ and $t_\alpha=\Delta_\alpha$, each minimal chain has a local charge-conjugation symmetry,
${\cal C}_{\alpha_1}H_\alpha(\phi_\alpha){\cal C}_{\alpha_1}^{-1}=H_\alpha(-\phi_\alpha)$, where 
the non-trivial action of the unitary operator ${\cal C}_{\alpha_1}$ is defined by 
${\cal C}_{\alpha_1}c_{\alpha_1}{\cal C}_{\alpha_1}^{-1}=c^\dagger_{\alpha_1} $, 
${\cal C}_{\alpha_1}c_{\alpha_2}{\cal C}_{\alpha_1}^{-1}=-e^{i\phi_\alpha}c_{\alpha_2}$,
${\cal C}_{\alpha_1}c_{\beta_i}{\cal C}_{\alpha_1}^{-1}=c_{\beta_i}$ $(\beta\neq \alpha)$. Note that ${\cal C}_{\alpha_1}$ switches only an electron at site 1 on chain $\alpha$ with its antiparticle. We also have a similar local charge-conjugation symmetry ${\cal C}_{\alpha_2}$, switching only an electron at site 2 with its antiparticle.
We can partially keep the symmetries even for the junctions in Eq.~(\ref{MKCH}).
For instance, by performing ${\cal C}_{{\rm L}_2}$, ${\cal C}_{{\rm M}_1}$, and an additional gauge rotation, we have  
${\cal C}_{\rm LM}H(\phi_{\rm L}, \phi_{\rm R}){\cal C}^{-1}_{\rm LM}=H(-\phi_{\rm L}, \phi_{\rm R})
$, where the unitary operator ${\cal C}_{\rm LM}$ acts as
${\cal C}_{\rm LM}c_{{\rm L}_1}{\cal C}^{-1}_{\rm LM}=e^{i\phi_{\rm L}}c_{{\rm L}_1}$, ${\cal C}_{\rm LM}c_{{\rm L}_2}{\cal C}^{-1}_{\rm LM}=c^\dagger_{{\rm L}_2}$,
${\cal C}_{\rm LM}c_{{\rm M}_1}{\cal C}_{\rm LM}^{-1}=-c^\dagger_{{\rm M}_1}$, ${\cal C}_{\rm LM}c_{{\rm M}_2}{\cal C}_{\rm LM}^{-1}=c_{{\rm M}_2}$, and ${\cal C}_{\rm LM}c_{{\rm R}_i}{\cal C}_{\rm LM}^{-1}=c_{{\rm R}_i}$. 
We also have the relation
${\cal C}_{\rm MR}H(\phi_{\rm L}, \phi_{\rm R}){\cal C}^{-1}_{\rm MR}=H(\phi_{\rm L}, -\phi_{\rm R})$
with an unitary operator ${\cal C}_{\rm MR}$ constructed similarly.
These relations provide the same constraint on the quasi-particle spectrum as local inversion, $E_n(\phi_{\rm L},\phi_{\rm R})=E_n(\pm \phi_{\rm L},\pm\phi_{\rm R})$, and thus, we need to break these local charge-conjugation symmetries,  in addition to local TRS, to obtain the Josephson diode effect. 
For this purpose, we consider $\Delta_{\rm M}\neq t_{\rm M}$ below,  which breaks both ${\cal C}_{\rm LM}$ and ${\cal C}_{\rm MR}$.

\begin{figure}[!t]
\centering
\includegraphics[width=0.49\textwidth]{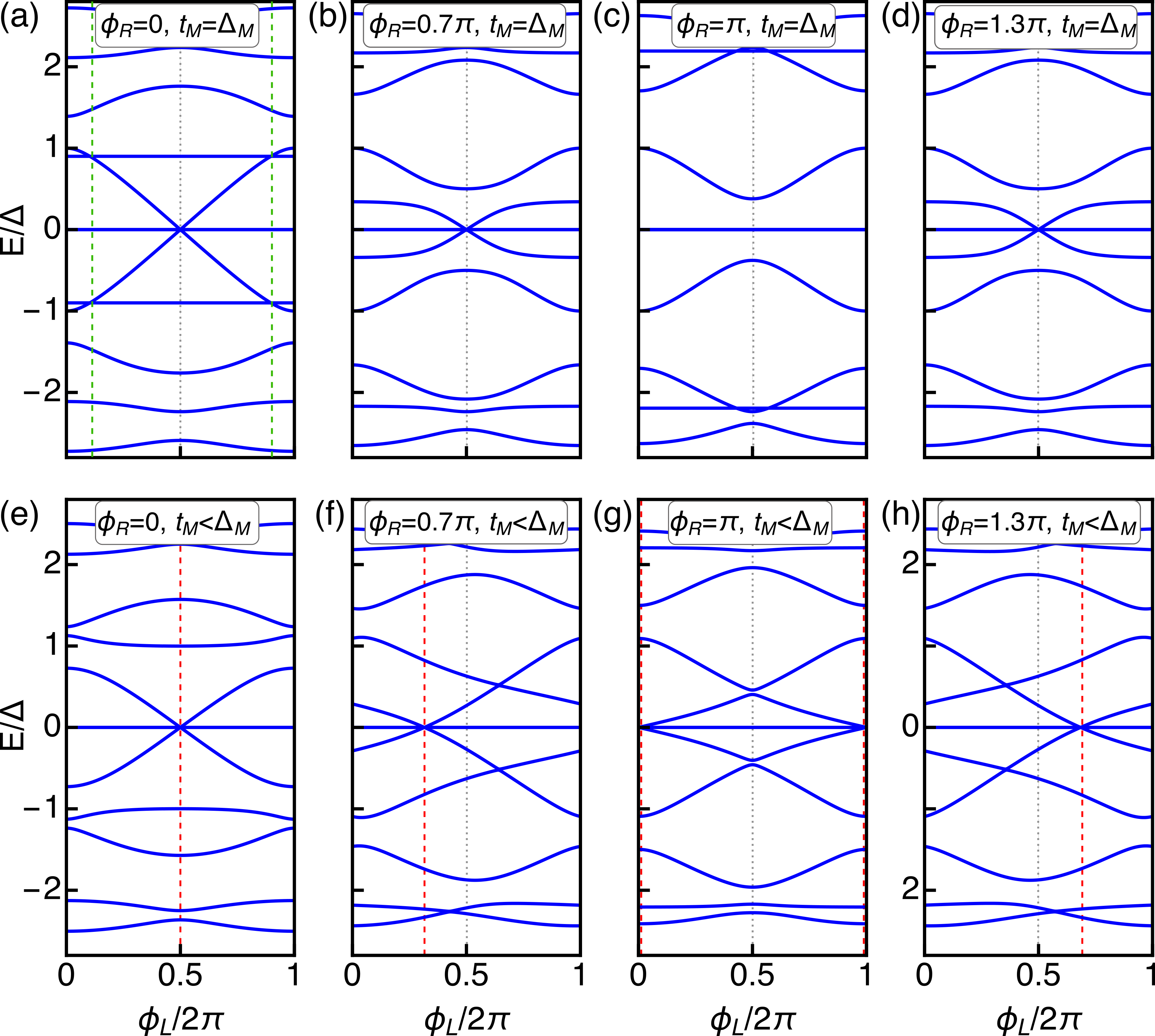}
\caption{Energy spectrum as a function of $\phi_{\rm L}$ at distinct $\phi_{\rm R}$ for $t_{\rm M}=\Delta_{\rm M}$ (a-d) and $t_{\rm M}<\Delta_{\rm M}$ (e-h). The vertical dotted line marks $\phi_{\rm L}=\pi$, while the red vertical dashed line marks  $\phi_{\rm L}$ at which the dispersing lowest ABSs reach zero energy. Parameters: $\varepsilon_{\alpha}=0$, $t_{\rm L,R}=\Delta_{\rm L,R}\equiv\Delta=1$, $\tau_{\rm L}=1$, $\tau_{\rm R}=0.9$.}
\label{Fig2} 
\end{figure}

\section{Andreev bound states and Andreev molecules}
Having identified the role of the superconducting phase differences $\phi_{\rm L,R}$ on the system symmetries, here we explore how they determine the behavior of the Andreev energy spectrum. This is shown in Fig.\,\ref{Fig2}(a-d) as a function of $\phi_{\rm L}$ when ECT and CAR  in the middle chain are the same ($t_{\rm M}=\Delta_{\rm M}$) for distinct values of $\phi_{\rm R}$ at $\Delta_{\rm L,R}=t_{\rm L,R}$, while in Fig.\,\ref{Fig2}(e-h) for  ECT weaker than CAR. Since $\Delta_{\rm L,R}=t_{\rm L,R}$, the left (right) chains host zero-energy dispersionless Majorana flat bands located in their first (second) QD, which remain robust against $\phi_{\rm L,R}$ [Fig.\,\ref{Fig2}]. Depending on the ratio between CAR and ECT in the middle chain, the Andreev spectrum exhibits a strong dependence on $\phi_{\rm L,R}$. For $t_{\rm M}=\Delta_{\rm M}$ at $\phi_{\rm R}=0$ [Fig.\,\ref{Fig2}(a)],  a pair of dispersionless Andreev bound states (ABSs) appear at energies close to $\pm\Delta$  and intersect a pair of dispersing  ABSs near $\phi_{\rm L}=0,2\pi$. See green vertical lines. These dispersionless ABSs originate from the right junction between the middle and right chains, and, as $\phi_{\rm R}$ takes finite values, they hybridize with the dispersing ABSs formed in the left junction between the left and middle chains [Figs.\,\ref{Fig2}(a-d)]. The dispersing hybridized ABSs, which can be seen as Andreev molecules studied before in conventional superconductors \cite{Pillet_2019,PhysRevResearch.1.033004,Su_2017,10.21468/SciPostPhysCore.2.2.009,PhysRevB.101.174506,matsuo2023phase,matsuo2023phase2,Coraiola_2023,PRXQuantum.5.020340}, reach    zero energy at $\phi_{\rm L}=\pi$ (gray vertical line) and become flat bands with $\phi_{\rm L}$ at $\phi_{\rm R}=\pi$ [Fig.\,\ref{Fig2}(c)], always maintaining a symmetric profile about $\phi_{\rm L}=\pi$. 

When $t_{\rm M}\neq\Delta_{\rm M}$, a nonzero $\phi_{\rm R}$   modifies the phase dependence of the spectrum and, notably, it also makes the spectrum  asymmetric about  $\phi_{\rm L}=\pi$ (vertical dashed red line), provided $\phi_{\rm R}\neq2\pi n$ with $n\in\mathbb{Z}$; see Fig.\,\ref{Fig2}(e-h) and Fig.\,\ref{Fig1EM} in 
Appendix  \ref{AppendixA}. This asymmetry moves the zero-energy crossing of the dispersing hybridized ABSs to $\phi_{\rm L}\neq\pi$, which occurs below (above) $\phi_{\rm L}=\pi$ when $\phi_{\rm R}<\pi$ ($\phi_{\rm R}>\pi$) for $t_{\rm M}<\Delta_{\rm M}$ [Fig.\,\ref{Fig2}(e-h)]. The asymmetric spectrum turns out to be gapless over $\phi_{\rm L,R}\in(0,2\pi)$ and remains for $t_{\rm M}>\Delta_{\rm M}$ but with the zero-energy crossing at  $\phi_{\rm L}<\pi$ ($\phi_{\rm L}>\pi$) when $\phi_{\rm R}>\pi$ ($\phi_{\rm R}<\pi$) in the latter case [Fig.\,\ref{Fig1EM}]. A consequence of the asymmetric spectrum under $\phi_{\rm L,R}\neq0$ and $\Delta_{\rm M}\neq t_{\rm M}$ is that the free energy develops minima away from $\phi_{\rm L}=2\pi n$, unlike conventional Josephson junctions \cite{RevModPhys.76.411,sauls2018andreev}; (See Appendix  \ref{AppendixA}). Thus, the imbalance between CAR and ECT in the middle Kitaev chain induces an asymmetric Andreev spectrum at nonzero superconducting phases across the left/right junctions,  in line  with breaking local TRS and local charge-conjugation symmetry. From an experimental point of view, we note that the  ratio between $\Delta_{\rm M}$ and $t_{\rm M}$ can be controlled e. g., by gating the middle chain as done in Ref.\,\cite{dvir2023realization} for tuning CAR and ECT processes.


\begin{figure}[!t]
\centering
\includegraphics[width=0.49\textwidth]{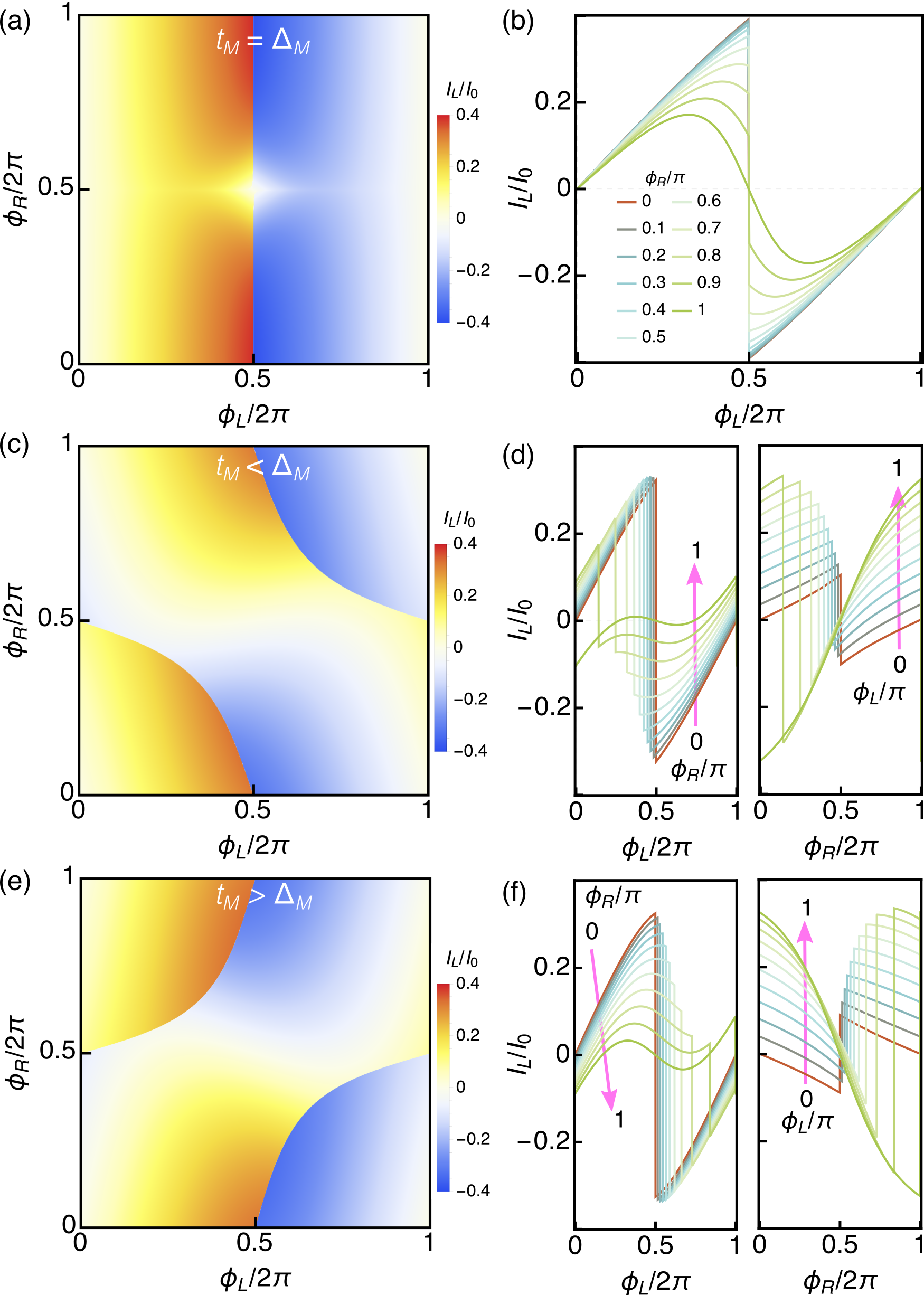}
\caption{(a,c,e) Supercurrent as a function of $\phi_{\rm L}$ and $\phi_{\rm R}$ for $t_{\rm M}<\Delta_{\rm M}$ (a), $t_{\rm M}=\Delta_{\rm M}$ (c), and $t_{\rm M}>\Delta_{\rm M}$ (e). (b,d,f) Line cuts as a function of $\phi_{\rm L (R)}$ for distinct  $\phi_{\rm R (L)}$. In (d,f), the magenta arrow indicates   variations of $\phi_{\rm L,R}$ within $(0,\pi)$ in steps of $0.1\pi$.
Parameters: $I_{0}=e\Delta/\hbar$, the rest as in Fig.\,\ref{Fig2}.}
\label{Fig3} 
\end{figure}

\section{Local and nonlocal Josephson effect}
A direct consequence of $\phi_{\rm L,R}\neq0$ and the imbalance between CAR and ECT in the middle chain is the phase-biased Josephson effect, which we explore here. For this purpose, we numerically obtain the phase-biased Josephson current flowing across the left junction as \cite{Beenakker:92,zagoskin,tanakaReview2024} $I_{\rm L}(\phi_{\rm L},\phi_{\rm R})=(e/\hbar)\sum_{n<0}dE_{n}(\phi_{\rm L},\phi_{\rm R})/d\phi_{\rm L}$ \footnote{Similarly, we can obtain the supercurrent flowing across the right junction as $I_{\rm R}(\phi_{\rm L},\phi_{\rm R})=(e/\hbar)\sum_{n<0}dE_{n}(\phi_{\rm L},\phi_{\rm R})/d\phi_{\rm R}$. We have verified that $I_{\rm L,R}(\phi_{\rm L},\phi_{\rm R})$ can also be obtained within a Green's function approach \cite{san2013multiple,zagoskin}, but obtaining them from the spectrum offers a direct way to identify the contributing states \cite{furusaki1991dc,Furusaki_1999,kashiwaya2000tunnelling,PhysRevB.96.205425,cayao2018andreev,PhysRevLett.123.117001,PhysRevB.104.L020501}.}. To visualize the supercurrent, in Figs.\,\ref{Fig3}(a,c,e) we show $I_{\rm L}$ as a function of $\phi_{\rm L}$ and $\phi_{\rm R}$ for distinct cases when CAR in the middle chain is equal (a), weaker (c), and stronger (e) than ECT in the same chain.  In Figs.\,\ref{Fig3}(b,d,f) we show $I_{\rm L}$ as a function of $\phi_{\rm L}$ for distinct $\phi_{\rm R}$, while   Figs.\,\ref{Fig3}(d,f) also show  $I_{\rm L}$ as a function of $\phi_{\rm R}$ for distinct $\phi_{\rm L}$. The first observation is that  $I_{\rm L}$ in Figs.\,\ref{Fig3} strongly depends on $\phi_{\rm L,R}$ and the ratio between ECT ($t_{\rm M}$) and CAR  ($\Delta_{\rm M}$) in the middle region, a behavior that originates from the ABSs in Figs.\,\ref{Fig2}. When  $\Delta_{\rm M}=t_{\rm M}$,  the supercurrent $I_{\rm L}$ is   $2\pi$-periodic and symmetric under $\phi_{\rm R}\neq0$ [Figs.\,\ref{Fig3}(a,b)], having $I_{\rm L}(\phi_{\rm L}=n\pi)=0$ for any $\phi_{\rm R}$ and    equal maxima/minima that reduce for $\phi_{\rm R}\neq0,2\pi$; the free energy in Fig.\,\ref{Fig2EM}(a) of Appendix \ref{AppendixB} has a minimum at $\phi_{\rm L}$ as for $0$-junctions \cite{RevModPhys.76.411,RevModPhys.77.935}. Also, $I_{\rm L}$  exhibits a sawtooth profile at $\phi_{\rm L}=\pi$ for $\phi_{\rm R}\neq \pi$  due to the zero-energy crossing of the dispersing ABSs [Figs.\,\ref{Fig2}(a-d)], while it smoothens out at  $\phi_{\rm R}=\pi$   because the dispersing ABSs become flat   [Figs.\,\ref{Fig2}(c)].

At $\Delta_{\rm M}\neq t_{\rm M}$, the supercurrent across the left junction $I_{\rm L}$ develops an intriguing anomalous behavior that varies with $\phi_{\rm L,R}\neq0$. See Figs.\,\ref{Fig3}(c-f).   The most notorious feature is that $\phi_{\rm R}\neq n\pi$ gives rise to $I_{\rm L}(\phi_{\rm L}=n\pi)\neq0$ with $n\in\mathbb{Z}$ and it is not symmetric about $\phi_{\rm L}=\pi$ anymore, which are consequences of the asymmetric Andreev spectrum shown in Figs.\,\ref{Fig2}(e-h).
This implies that, even at $\phi_{\rm L}=0$, a supercurrent across the left junction $I_{\rm L}$ emerges and is controlled by the superconducting phase difference across the right junction $\phi_{\rm R}$,  signalling  the so-called nonlocal JE which was initially predicted in systems with conventional superconductors \cite{Pillet_2019}. This nonlocal JE requires multiple terminal and does not appear in a two-terminal Josephson junction. The sawtooth profile of $I_{\rm L}$ also occurs away from $\phi_{\rm L}=\pi$ [Figs.\,\ref{Fig3}(d,f)] due to the shifted zero-energy crossing in the ABSs when  $t_{\rm M}\neq\Delta_{\rm M}$. See   Figs.\,\ref{Fig2}(e-h) and Figs.\,\ref{Fig1EM}. It is worth noting that, having $I_{\rm L}(\phi_{\rm L}=0,\phi_{\rm R}\neq n\pi)\neq0$ in Figs.\,\ref{Fig3}(c-f)  also means that  a $\phi_{\rm L,0}$-Josephson junction is realized at $\Delta_{\rm M}\neq t_{\rm M}$ \cite{RevModPhys.76.411,RevModPhys.77.935,PhysRevLett.101.107005}. At $\phi_{\rm L}=0$, the current as a function of $\phi_{\rm R}$ can develop a $\pi$-shift for $t_{\rm M}>\Delta_{\rm M}$ [Figs.\,\ref{Fig3}(e,f)], leading to a $\pi$-junction \cite{RevModPhys.77.935,PhysRevB.111.184515}. Moreover, there are also   signals of a $\phi$-junction behavior \cite{PhysRevB.57.R3221,PhysRevB.67.220504,PhysRevLett.133.226002,PhysRevB.111.064502}, where the current $I_{\rm L}$ passes through zero neither at $\phi_{\rm L}=0,2\pi$ nor $\phi_{\rm L}=\pi$ [Figs.\,\ref{Fig3}(d,f)].  Yet another important effect of $\phi_{\rm L,R}\neq0$ and $t_{\rm M}\neq\Delta_{\rm M}$ is that the maximum positive and negative values of $I_{\rm L}$ over  $\phi_{\rm L}\in(0,2\pi)$ are not the same [Figs.\,\ref{Fig3}(c-f)], producing a nonreciprocity between the maximum currents that can flow across the left junction. As a result, a Josephson setup with three laterally coupled minimal Kitaev chains [Eq.\,(\ref{MKCH})] can realize the nonlocal Josephson effect and nonlocal nonreciprocal Josephson transport only when ECT and CAR in the middle chain are distinct.

\begin{figure}[!t]
\centering
\includegraphics[width=0.49\textwidth]{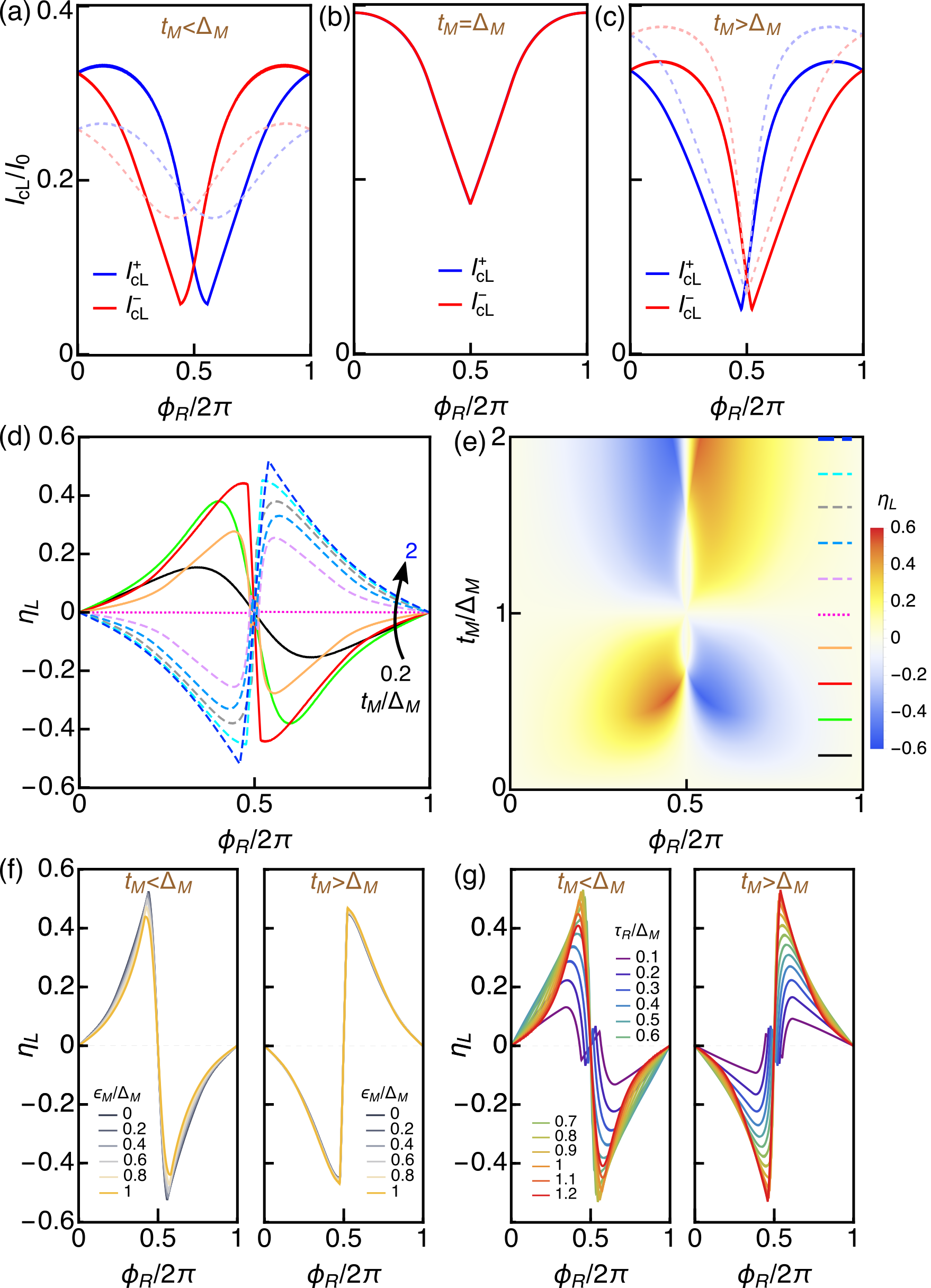}
\caption{(a-c) Critical currents $I_{\rm cL}^{\pm}$ as functions of $\phi_{\rm R}$ for $t_{\rm M}<\Delta_{\rm M}$ (a), $t_{\rm M}=\Delta_{\rm M}$ (b), and $t_{\rm M}>\Delta_{\rm M}$ (c). The dashed and solid lines in (a) correspond to $t_{\rm M}=0.2\Delta_{\rm M}$ and $t_{\rm M}=0.5\Delta_{\rm M}$, while to $t_{\rm M}=1.5\Delta_{\rm M}$ and $t_{\rm M}=1.8\Delta_{\rm M}$ in (c). (d,e) Diode's quality factor $\eta_{\rm L}$ as a function of $\phi_{\rm R}$ for distinct $t_{\rm M}$ in steps of $0.2\Delta_{\rm M}$, marked by colored bars in (e). (f)  $\eta_{\rm L}$ as a function of $\phi_{\rm R}$ for distinct $\varepsilon_{\rm M}$ for $t_{\rm M}<\Delta_{\rm M}$ and $t_{\rm M}>\Delta_{\rm M}$. (g) The same as in (f) but for distinct $\tau_{\rm R}$. Parameters: $I_{0}=e\Delta/\hbar$, the rest as in Fig.\,\ref{Fig2}.}
\label{Fig4} 
\end{figure}

\section{Critical currents and nonlocal Josephson diode effect}
To further explore the nonlocal nonreciprocal Josephson transport, we look at the critical currents across the left junction $I_{\rm cL}^{\pm}(\phi_{\rm R})={\rm max}_{\phi_{\rm L}}[\pm I_{\rm L}(\phi_{\rm L},\phi_{\rm R})]$, where $I_{\rm L}(\phi_{\rm L},\phi_{\rm R})$ is the current flowing across the left junction discussed in the previous section.  Figs.\,\ref{Fig4}(a-c) show the critical currents $I_{\rm cL}^{\pm}(\phi_{\rm R})$   as a function of $\phi_{\rm R}$ when ECT in the middle chain is weaker, equal, and stronger than CAR in the same chain. At $t_{\rm M}=\Delta_{\rm M}$, when CAR and ECT in the middle chain are equal,  the positive and negative critical currents are equal ($I_{\rm cL}^{+}(\phi_{\rm R})=I_{\rm cL}^{-}(\phi_{\rm R})$) and develop a minimum at $\phi_{\rm R}=\pi$ [Fig.\,\ref{Fig4}(b)], leading to the absence of the nonlocal Josephson diode effect. Interestingly, for $t_{\rm M}<\Delta_{\rm M}$, the minima of $I_{\rm cL}^{+}(\phi_{\rm R})$ and $I_{\rm cL}^{-}(\phi_{\rm R})$ are  shifted to values below and above $\phi_{\rm R}=\pi$, respectively, giving rise to nonreciprocal phase-biased Josephson currents ($I_{\rm cL}^{+}(\phi_{\rm R})\neq I_{\rm cL}^{-}(\phi_{\rm R})$) flowing across the left junction and fully controlled by $\phi_{\rm R}\neq n\pi$. For $t_{\rm M}>\Delta_{\rm M}$, the shifts in the critical currents is the opposite as for $t_{\rm M}<\Delta_{\rm M}$ but $I_{\rm cL}^{+}(\phi_{\rm R})\neq I_{\rm cL}^{-}(\phi_{\rm R})$ remains. Having $I_{\rm cL}^{+}(\phi_{\rm R})\neq I_{\rm cL}^{-}(\phi_{\rm R})$ defines the nonlocal Josephson diode effect, which requires an imbalance between CAR and ECT in the middle chain and also $\phi_{\rm R}\neq n\pi$.   These conditions are consistent with the asymmetric Andreev spectrum [Fig.\,\ref{Fig2}] and   with the breaking of local TRS and local charge-conjugation symmetry discussed below Eq.\,(\ref{MKCH}), implying that they are crucial for the nonlocal Josephson diode effect in minimal Kitaev chains.

To characterize the nonlocal diode effect, we analyze its efficiency by studying the quality factor $\eta_{\rm L}(\phi_{\rm R})=[I_{\rm cL}^{+}(\phi_{\rm R})-I_{\rm cL}^{-}(\phi_{\rm R})]/[I_{\rm cL}^{+}(\phi_{\rm R})+I_{\rm cL}^{-}(\phi_{\rm R})]$, where $I_{\rm cL}^{\pm}(\phi_{\rm R})$ are the critical currents across the left junction analyzed in the previous paragraph.  We show $\eta_{\rm L}(\phi_{\rm R})$ in  Fig. \,\ref{Fig4}(d,e) as a function of $\phi_{\rm R}$ and $t_{\rm M}$, while in Fig. \,\ref{Fig4}(f,g) we inspect the stability of the efficiencies when varying the onsite energies and tunneling across the junction since these conditions are very likely unavoidable in a realistic scenario. In Appendix  \ref{AppendixC}, we discuss the robustness of  $\eta_{\rm L}(\phi_{\rm R})$ against variations of the onsite energies $\varepsilon_{\alpha}$, see Fig.\,\ref{Fig3EM}. In Fig. \,\ref{Fig4}(d,e), we observe that a nonzero quality factor appears when $t_{\rm M}\neq\Delta_{\rm M}$ and $\phi_{\rm R}\neq n\pi$, with $n\in\mathbb{Z}$, and it is enhanced  just below and above $\phi_{\rm R}=\pi$  achieving values above $0.5$. While $t_{\rm M}/\Delta_{\rm M}<1$ promotes a quality factor having a sine-line profile with $\phi_{\rm R}$ with a fast sign change across $\phi_{\rm R}=\pi$, the quality factors change sign for $t_{\rm M}/\Delta_{\rm M}>1$, see Fig. \,\ref{Fig4}(d,e). This implies that the diode's polarity and large efficiencies can be fully controlled nonlocally by $\phi_{\rm R}$ and is thus expected to be more tunable than regular two-terminal Josephson diodes \cite{PhysRevLett.99.067004,PhysRevB.93.174502,PhysRevB.92.035428,PhysRevB.103.245302,PhysRevB.106.214524,PhysRevLett.129.267702,davydova2022universal,PhysRevB.107.245415,PhysRevB.108.214520,PhysRevB.108.054522,PhysRevB.109.L081405,PhysRevB.110.014519,10.1063/5.0210660,PhysRevApplied.21.054057,PhysRevLett.131.096001,10.1063/5.0211491,PhysRevResearch.6.023011,PhysRevB.109.174511,79tj-c3y4}.  By  direct inspection of Fig. \,\ref{Fig4}(e), one concludes that, for $t_{\rm M}<\Delta_{\rm M}$,  the largest efficiencies are achieved for $t_{\rm M}\in(0.3-0.7)\Delta_{\rm M}$ and   $0.3\pi<\phi_{\rm R}<\pi$ ($\pi<\phi_{\rm R}<0.7\pi$). The large efficiencies of the nonlocal Josephson diode effect are robust against variations of the onsite energies of the middle chain, as demonstrated in  Figs.\,\ref{Fig4}(f). Strong variations of the onsite energies in the left and right chains, however, have a detrimental effect, while $\eta_{\rm L}(\phi_{\rm R})$  remains robust against weak detuning, see Fig.\,\ref{Fig3EM}. Moreover, having large tunneling amplitudes favours the large efficiency values of the nonlocal Josephson diode  [Fig.\,\ref{Fig4}(g)]. We have also verified that slight deviations from the equal CAR and ECT regime in the left /right chains weakly affects the nonlocal Josephson diode effect.

\section{Conclusions}
In conclusion, we have demonstrated the nonlocal Josephson effect and the realization of a nonlocal Josephson diode in a system formed by three laterally coupled minimal Kitaev chains. In particular, we found that when the middle chain exhibits an imbalance between crossed Andreev reflections and electron cotunneling, the phase-dependent Andreev spectrum is asymmetric and can be controlled by the phase difference across the right junction. We have then shown that this asymmetric Andreev spectrum, containing hybridized Andreev bound states from both junctions, produces a supercurrent across the left junction that is controllable by the phase difference across the right junction, hence defining the nonlocal Josephson effect.  Notably, we have discovered that these nonlocal Josephson effects enable distinct positive and negative critical currents, giving rise to a nonlocal Josephson diode effect whose polarity is highly tunable and can achieve efficiencies above $50\%$.  Given the recent advances on nonlocal Josephson effect \cite{strambini2016omega,draelos2019supercurrent,PhysRevX.10.031051,graziano2022selective,Matsuo_2022,matsuo2023phase,Haxell_2023} and minimal Kitaev chains \cite{dvir2023realization,bordin2023crossed,PhysRevX.13.031031,zatelli2023robustpoorMajo}, our findings are very likely within experimental reach. In this regard, we point out that Ref.\,\cite{dvir2023realization} has already experimentally demonstrated the control of the ratio between crossed Andreev reflections and electron cotunneling, which, applied to the middle chain in our setup, shows that the realization of the nonlocal Josephson diode effect with a highly tunable polarity in minimal Kitaev chains is a feasible idea. While our work makes us wonder about emergent effects in minimal Kitaev chains, an immediate question also involves the application of the nonlocal Josephson diode concept in longer Kitaev chains, which will be the focus of our future research.

\begin{acknowledgments}
 J. C. acknowledges financial support from the Swedish Research Council (Vetenskapsr{\aa}det Grant No. 2021-04121) and from the Olle Engkvist Foundation (Grant No.  243-1026). M. S. was supported by JST CREST Grant No. JPMJCR19T2 and JSPS KAKENHI Grant Nos. JP24K00569 and JP25H01250. The computations were enabled by resources provided by the National Academic Infrastructure for Supercomputing in Sweden (NAISS), partially funded by the Swedish Research Council through Grant Agreement No. 2022-06725.
\end{acknowledgments}	 
  

  \appendix

\section{Andreev bound states when ECT is stronger than CAR in the middle chain}
\label{AppendixA}
\begin{figure}[!t]
\centering
\includegraphics[width=0.49\textwidth]{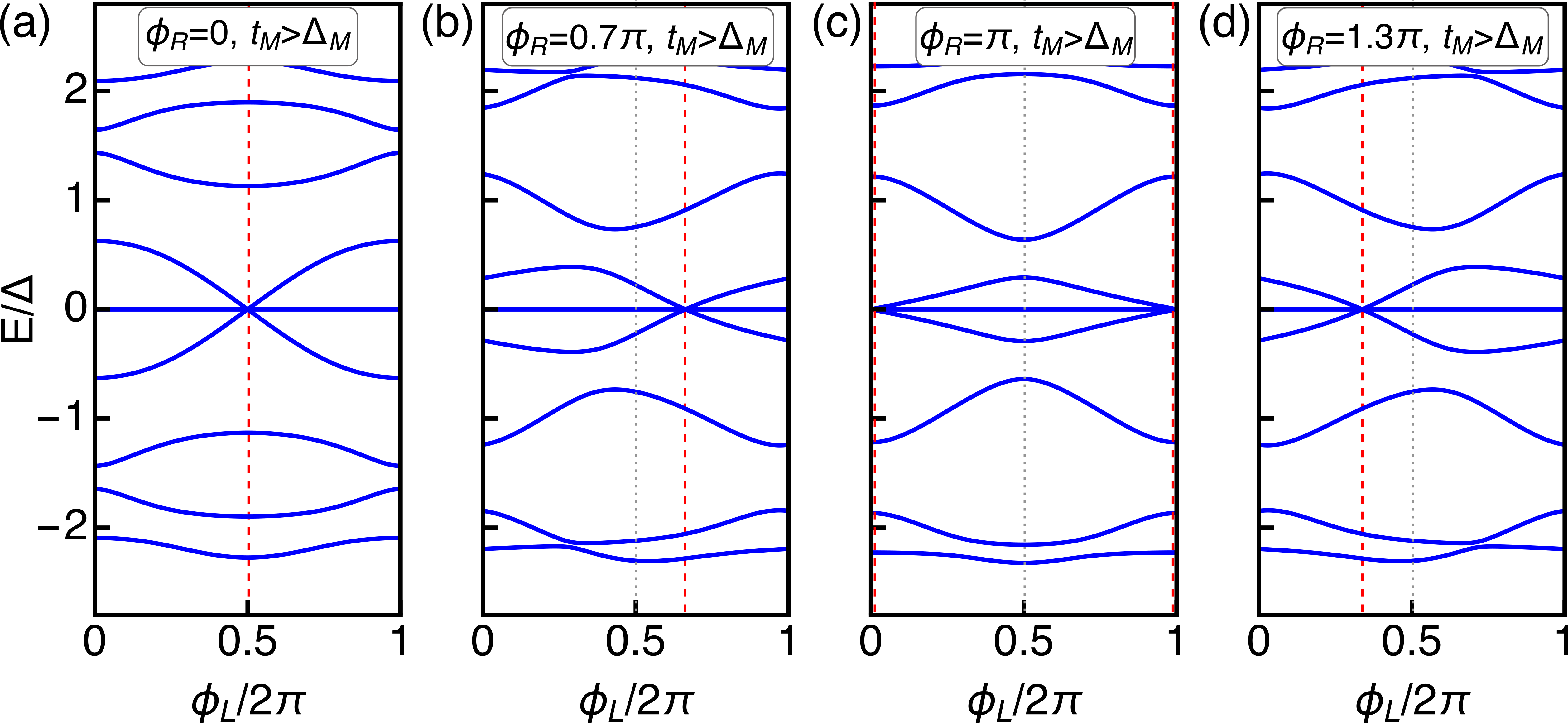}
\caption{Energy spectrum as a function of $\phi_{\rm L}$ at distinct values of $\phi_{\rm R}$ for   $t_{\rm M}=1.8\Delta_{\rm M}$. The vertical dotted line marks $\phi_{\rm L}=\pi$, while the red vertical dashed line marks the value of $\phi_{\rm L}$ at which the dispersing lowest ABSs reach zero energy. Parameters: $\varepsilon_{\alpha}=0$, $t_{\rm L,R}=\Delta_{\alpha}\equiv\Delta=1$, $\tau_{\rm L}=1$, $\tau_{\rm R}=0.9$.}
\label{Fig1EM} 
\end{figure}

In Fig.\,\ref{Fig2} we looked at the Andreev spectrum when ECT is equal to or weaker than CAR in the middle chain. In this part,  Fig.\,\ref{Fig1EM}  shows the Andreev spectrum as a function of $\phi_{\rm L}$ for distinct $\phi_{\rm R}$ when ECT is stronger than CAR, namely, $t_{\rm M}>\Delta_{\rm M}$. The Andreev spectrum is symmetric when $\phi_{\rm R}=0$ [Fig.\,\ref{Fig1EM}(a)], developing a phase dependence similar to the case with $t_{\rm M}<\Delta_{\rm M}$ in Fig.\,\ref{Fig2}(a). Because we chose $\Delta_{\rm L,R}=t_{\rm L,R}$ and $\varepsilon_{\alpha}$, the spectrum contains two dispersionless levels at zero energy, which represent the two poor man's Majorana modes located at the very outer QDs. Two extra Andreev bound states reach zero energy at $\phi=\pi$ marked by a vertical dashed red line, and those are located in the middle of the junction. However, deviations from these parameters do not change the symmetric nature of the spectrum in Fig.\,\ref{Fig1EM}(a) unless $\phi_{\rm R}\neq n\pi$. This is demonstrated in   Fig.\,\ref{Fig1EM}(b-d), where a  $\phi_{\rm R}\neq n\pi$ makes the entire Andreev spectrum asymmetric with respect to $\phi_{\rm L}=\pi$, see vertical dashed gray line. The zero-energy crossing seen at $\phi_{\rm L}=0$ in Fig.\,\ref{Fig1EM}(a) shifts to $\phi_{\rm L}>\pi$ when $\phi_{\rm R}<\pi$ [Fig.\,\ref{Fig1EM}(b)], while to $\phi_{\rm L}<\pi$  for $\phi_{\rm R}>\pi$ [Fig.\,\ref{Fig1EM}(d)], see vertical dashed red lines. At $\phi_{\rm R}=\pi$, the Andreev specturm is symmetric, the pair of dispersionless levels remain, but the first excited levels now reach zero energy at $\phi_{\rm L}=2n\pi$ while leaving a large zero energy splitting at $\phi_{\rm L}=n\pi$, Fig.\,\ref{Fig1EM}(c). In conclusion, the asymmetry in the Andreev spectrum is controlled by $\phi_{\rm R}$ and $\Delta_{\rm M}\neq t_{\rm M}$, in line with the  breaking local time-reversal and local charge-conjugation symmetries discussed in the main text.

\section{Free energy of the Josephson setup}
\label{AppendixB}
\begin{figure}[!t]
\centering
\includegraphics[width=0.49\textwidth]{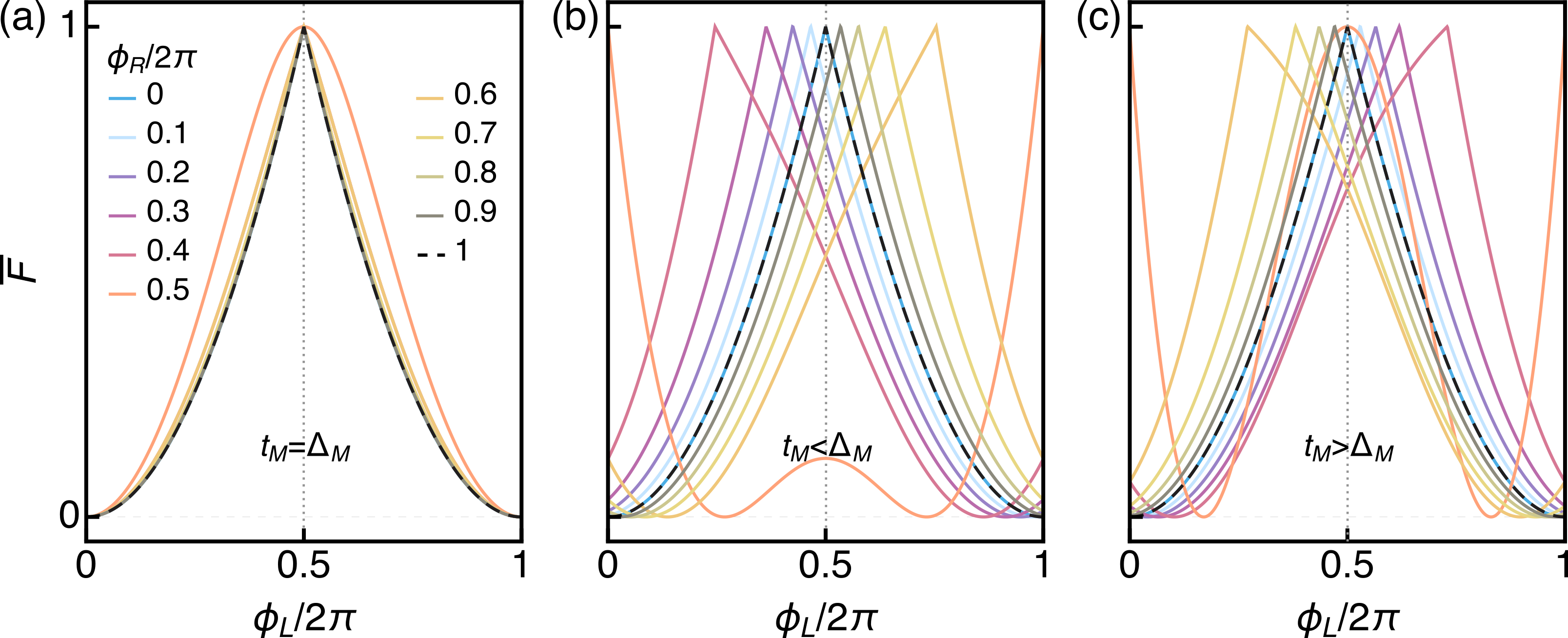}
\caption{Free energy $\bar{F}$ as a  function of $\phi_{\rm L}$ at distinct values of $\phi_{\rm R}$ for    $t_{\rm M}=\Delta_{\rm M}$,  $t_{\rm M}=0.5\Delta_{\rm M}$, and $t_{\rm M}=1.8\Delta_{\rm M}$. Here, $\bar{F}$  is defined as $\bar{F}=(F-F_{\rm min})/(F_{\rm max}-F_{\rm min})$, where $F$ is the free energy and $F_{\rm max(min)}$ is  the maximum (minimum) free energy value over a period of $\phi_{\rm L}$. Parameters: $\varepsilon_{\alpha}=0$, $t_{\rm L,R}=\Delta_{\alpha}\equiv\Delta=1$, $\tau_{\rm L}=1$, $\tau_{\rm R}=0.9$.}
\label{Fig2EM} 
\end{figure}

The asymmetric Andreev spectrum showed in the previous section has a direct effect on the free energy, which we discuss in this part.  We define a normalized free energy as $\bar{F}=(F-F_{\rm min})/(F_{\rm max}-F_{\rm min})$, where $F_{\rm max}={\rm max}_{\phi_{\rm L}}[F(\phi_{\rm L},\phi_{\rm R})]$ and  $F_{\rm min}={\rm min}_{\phi_{\rm L}}[F(\phi_{\rm L},\phi_{\rm R})]$, where the maximum and minimum are obtained over a period of $2\pi$. In Fig.\,\ref{Fig2EM}, we show $\bar{F}$ as a function of $\phi_{\rm L}$ for distinct values of $\phi_{\rm R}$ when ECT in the middle chain $t_{\rm M}$ is equal, weaker, and stronger than CAR in the same chain $\Delta_{\rm M}$. As anticipated, for $t_{\rm M}=\Delta_{\rm M}$ in Fig.\,\ref{Fig2EM}(a), the free energy $\bar{F}$ develops minima at multiples of $2n\pi$, with $n\in \mathbb{Z}$, irrespective of the values of $\phi_{\rm R}$. This signals a regular Josephson junction behavior. Interestingly, for $t_{\rm M}\neq\Delta_{\rm M}$ and $\phi_{\rm R}\neq0$, the free energy exhibits minima away from $\phi_{\rm L}=2n\pi$, see Fig.\,\ref{Fig2EM}(b,c). For instance, for $\phi_{\rm R}=\pi$ at $t_{\rm M}\neq\Delta_{\rm M}$, the free energy has two (double) minima away from $\phi_{\rm L}=0,\pi$, signaling the emergence of a $\phi$-Josephson junction. We note that $\phi$-Josephson junctions were initially predicted  in Josephson junctions with alternating critical current density \cite{PhysRevB.57.R3221}, in periodic arrays of $0$- and $\pi$-Josephson junctions \cite{PhysRevB.67.220504}, and more recently in altermagnetic Josephson junctions  \cite{PhysRevLett.133.226002,PhysRevB.111.064502,FukayaJPCM2025}. Furthermore, away from $\phi_{\rm R}=\pi$ at $t_{\rm M}\neq\Delta_{\rm M}$, the free energy is able to exhibit one minimum either close to $\phi_{\rm L}=0$  or $\phi_{\rm L}=2\pi$: on one hand, when $\pi<\phi_{\rm R}<2\pi$, the minimum is   close to $\phi_{\rm L}=0$  for   $t_{\rm M}<\Delta_{\rm M}$ but   close to $\phi_{\rm L}=2n\pi$ for  $t_{\rm M}>\Delta_{\rm M}$. On the other hand, when $0<\phi_{\rm R}<\pi$, the minimum is close to $\phi_{\rm L}=2\pi$ for  $t_{\rm M}<\Delta_{\rm M}$ but   close to $\phi_{\rm L}=0$ for  $t_{\rm M}>\Delta_{\rm M}$. The asymmetric Andreev spectrum and nonreciprocal (diode) Josephson transport discussed in the main text
is therefore tied to the  free energy having a single minimum away from $\phi_{\rm L}=2n\pi$.

\section{Efficiency of the nonlocal Josephson diode at finite onsite energies}
\label{AppendixC}
\begin{figure}[!t]
\centering
\includegraphics[width=0.49\textwidth]{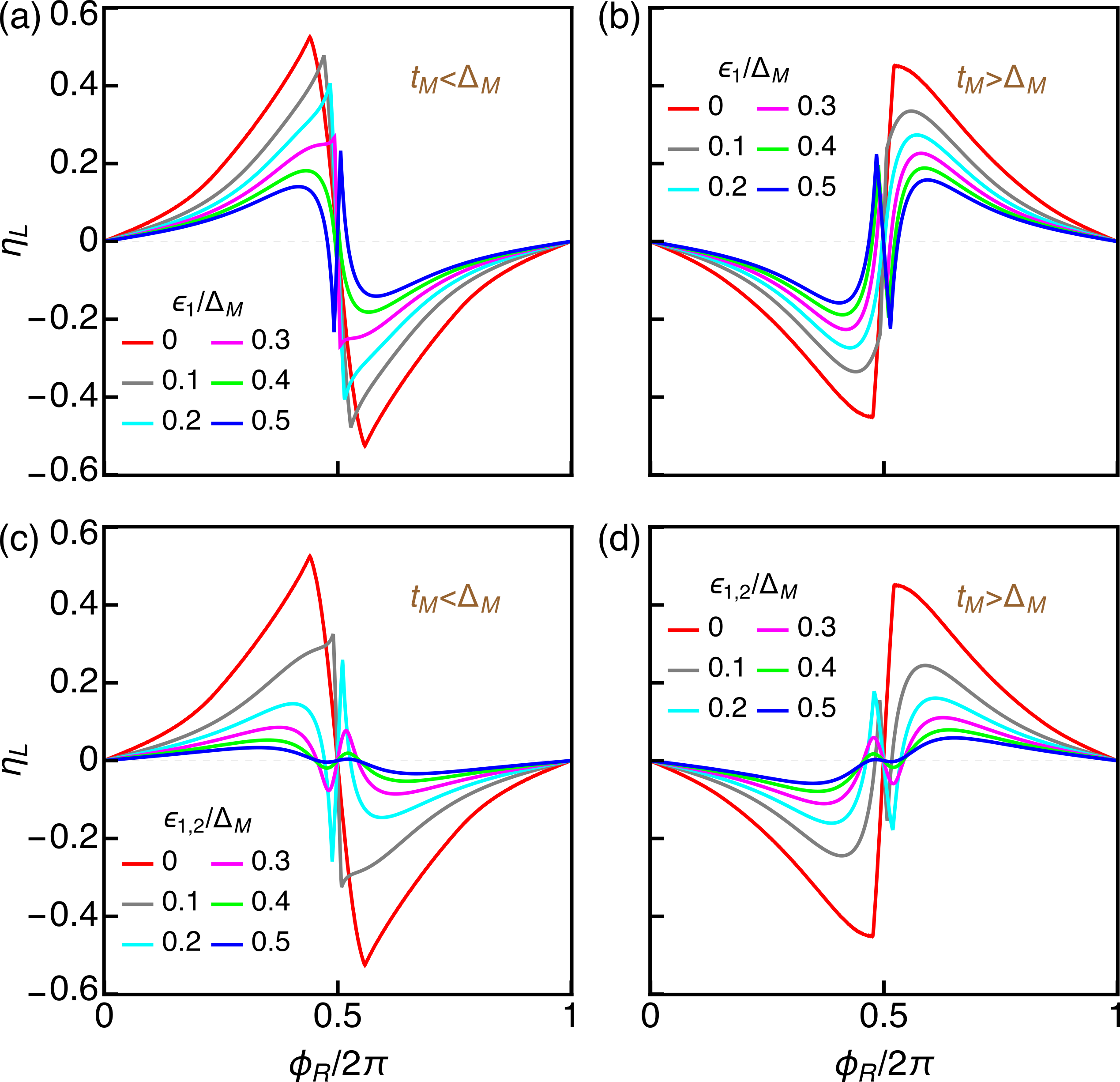}
\caption{(a,b) Diode's efficiency as a function of $\phi_{\rm R}$ for distinct values of  the onsite energy $\varepsilon_{\rm L_1,R_1}=\epsilon_{1}$ when $t_{\rm M}=0.5\Delta_{\rm M}$ (a) and    $t_{\rm M}=1.8\Delta_{\rm M}$ (a), in both cases at $\varepsilon_{\rm L_{2}(R_{2})}=0$. (c,d) The same as in (a,b) but both onsite energies     $\varepsilon_{\rm L_{1,2}}=\varepsilon_{\rm R_{1,2}}=\epsilon_{1,2}$. Parameters: $\varepsilon_{\rm M}=0$, $t_{\rm L,R}=\Delta_{\alpha}\equiv\Delta=1$, $\tau_{\rm L}=1$, $\tau_{\rm R}=0.9$.}
\label{Fig3EM} 
\end{figure}

To further support the robustness of the nonlocal Josephson diode effect presented in Fig.\,\ref{Fig4}, here we explore the sensitivity of the quality factor $\eta_{\rm L}$ under variations of the onsite energies $\varepsilon_{\alpha}$. This is presented in Fig.\,\ref{Fig3EM}(a,b), where we show $\eta_{\rm L}$ as a function of $\phi_{\rm R}$ when the onsite energies of the first QD of the left and right chains take finite values, while in  Fig.\,\ref{Fig3EM}(b,c) we consider the case when both onsite energies of the left and right chains are nonzero and equal. 
The quality factors exhibit a slightly greater robustness against variations of a single onsite energy [Fig.\,\ref{Fig3EM}(a,b)], achieving values close to $20\%$ for detuning values of  $\epsilon_{1}/\Delta=0.5$. When both QD energies of the outer chains equally large, the quality factors degrade faster than when a single QD has a nonzero energy; see Fig.\,\ref{Fig3EM}(c,d).

While these QD energies seem to be detrimental, it is worth noting that they also bring additional functionality for the nonlocal Josephson diode. To see this, we first notice that the polarity of the diode, reflected in the sign of $\eta_{\rm L}$, is highly controllable by the phase across the junction between the middle and right chains  $\phi_{\rm R}$. The diode's polarity is also determined by strength of ETC and CAR, as  observed in Fig.\,\ref{Fig3EM}. Notably, the nonzero values of the onsite energies in the outer chains have the capability to change the diode's polarity around $\phi_{\rm R}=\pi$. Depending on whether a single  $\epsilon_{1}$ or both $\epsilon_{1,2}$ energies are varied, the change in the diode's polarity can be fast or slow, respectively. Since the QD energies are expected to be tunable by means of gates \cite{dvir2023realization}, $\epsilon_{1,2}$ add important control of the nonlocal Josephson diode effect discussed here.

\bibliography{biblio}

\end{document}